\documentclass[runningheads]{llncs}
\usepackage{url}
\usepackage{enumerate}
\usepackage{url}
\usepackage{todonotes}
\usepackage{algorithm}
\usepackage{algpseudocode}
\usepackage{cancel}
\usepackage{graphicx}
\usepackage{subcaption}
\usepackage{comment}
\usepackage{amsmath,amssymb,amsfonts}
\usepackage{booktabs} 
\usepackage{multirow}
\usepackage[T1]{fontenc}        

\newcommand{\SV}[1]{#1}
\newcommand{\LV}[1]{}

\def \beq{\begin{eqnarray*}}
\def \eeq{\end{eqnarray*}}

\begin{document}

\title{A Predictive Framework for Base-$n$ Radix Sort Optimization}
\author{Atharv Pandey\inst{1}\orcidID{0009-0000-6456-0240}
\and \\ Lakshmanan Kuppusamy\inst{1}\orcidID{0000-0003-2358-905X}}
\authorrunning{A. Pandey and L. Kuppusamy}

\institute{School of Computer Science and Engineering, VIT University, \\
Vellore - 632~014, India. \email{klakshma@vit.ac.in} }  
\date{}
\maketitle

\begin{abstract}
Sorting is a foundational primitive of computer science and optimizations in sorting subroutines can cascade into significant performance gains for high-throughput systems.
In this paper, we analyze the inefficiencies of a non-comparison sorting algorithm, namely, {\it Base-$n$ Radix Sort} (BNRS), specifically the `zero padding' problem in skewed datasets. We develop an execution model, called, {\it Stable Partitioning - Least Significant Digit Radix Sort} (shortly, SP-LSD), an iterative least significant digit based pruning model designed to address this inefficiency.
Based on this development, we derive the {\it Radix Crossover Framework} (RCF), an analytic three-point decision framework. The framework is established on the precondition of non-negative integers, which enables the derivation of three critical boundaries. First, the {\it Asymptotic Crossover} ($k < n^{\log_2 n}$) defines when BNRS and SP-LSD can theoretically outperform the comparison sorting algorithms where $k$ is the maximum value and $n$ is the input size. Second, the {\it Round-feasibility Crossover} ($k > n^2$) defines when overhead cost of implemented model SP-LSD is amortized. Third, we derive {\it Pruning Crossover} parameterized by the ratio of random-access sorting cost to sequential partitioning cost. This model demonstrates that SP-LSD yields a net gain on skewed and uniform distributions over standard BNRS. The experimental results are consistent with the crossover boundaries, providing a deterministic roadmap for adaptive algorithm selection.
\end{abstract}

\section{Introduction}
The typical application of Radix sort uses base 10 but Base-$n$ Radix Sort
(shortly, BNRS) \cite{knuth98} uses base-$n$ instead of base 10, where $n$ is the input size. The working nature of BNRS is discussed in detail in the next section and a demonstration is provided in Appendix. BNRS suffers from `zero padding problem' for certain data distributions. More specifically, this is caused due to non-uniform effective bit widths which occurs when the input dataset has varying significant digit counts (i.e. magnitudes). The smaller magnitude values are treated by appending leading zeros (i.e., padded with zeros from left) to match the key length of the maximum element. The total count of rounds are defined by $\lfloor \log_n k \rfloor + 1$, where $k$ is the maximum element, the iteration of each round defines a log group i.e. the elements ranging from $n^{r-1}$ to $n^{r}-1$ where $r$ is the iteration of rounds. Depending on the input distribution, there would be execution of redundant passes performed for the input values that have a smaller effective width. This results in redundant computation which continues operations on padded keys for elements that have already been positioned correctly within their respective log groups in the previous rounds.  

Another non-comparison sorting that we discuss and benchmark in this paper is American Flag Sorting (AFS) \cite{ARS93}, which ia a variant of Radix sort. While Radix sort uses LSD approach, AFS uses MSD approach. See Appendix for details of this AFS sorting and its weaknesses: {\it cache inefficiency} and {\it fat bucket phenomenon}. The time and space complexity of used comparison and non-comparison sorting algorithms are given in Appendix, Table \ref{tab:complexity}. 

The following observation motivates the work for this paper. A closer analysis of the BNRS algorithm reveals that for common skewed data distributions, a large portion of elements - relative to each other in terms of the same $\log$ group, are effectively sorted after the respective $\log$ rounds end. As a result, our research is to optimize the BNRS for heavy-tailed skewed datasets. We contend this through adaptively sorting the unsorted portion, thereby reducing the actual number of operations to below the claimed bounds\footnote{It is a usual practice to ignore the base for logarithmic functions when denoted in Big $O$ notation. However, we would like to specify the base to signify that $n$ is the input size.} of $O(n \log_n k)$. 
\newcommand{\tcsort}{
The detailed time complexity some comparison and non-comparison sorting algorithms is given in Table \ref{tab:complexity}. 
\begin{table}[htbp]
\centering
\caption{Complexity analysis of known non-comparison sorting algorithms, where $n$ is input size, $k$ is maximum value, $d$ is number of rounds/digits and $b$ is radix/base for Radix Sort and number of buckets for Bucket Sort. }
\label{tab:complexity}
\begin{tabular}{|c|c|c|c|c|}
\hline
\textbf{Algorithm} & \textbf{Best Case} & \textbf{Average Case} & \textbf{Worst Case} & \textbf{Space} \\
\hline
Counting Sort & $O(n+k)$ & $O(n+k)$ & $O(n+k)$ & $O(n+k)$ \\
Radix Sort & $O(d \cdot (n+b))$ & $O(d \cdot (n+b))$ & $O(d \cdot (n+b))$ & $O(n+b)$ \\
Bucket Sort & $O(n+b)$ & $O(n+b)$ & $O(n^2)$ & $O(n+b)$ \\
\hline 
\end{tabular}
\end{table}
}

The primary contribution of our research is to introduce a new optimization approach for BNRS, called Stable Partition - Least Significant Digit Radix Sort (SP-LSD). In this introduced method, after the first iteration, the original input array is divided into two logical regions, `sorted' and `active' regions. The active region is processed every round and the size of it is gets reduced in the subsequent rounds. The decision of whether an element is sorted based on the condition $A[i] / base^{r-1} = 0$. To show that our proposed approach outperforms both the comparison sorts (e.g., Introsort) and non-comparison sorts (e.g., BNRS, AFS), we develop a {Radix Crossover Framework}, involving theoretical analysis and practical implementation containing a three point decision framework, namely {\it Aymptotic}, {Round-feasibility} and {\it Pruning} crossovers. Further, in our analysis, we demonstrate that when very large and skewed input is received, e.g., say sorting input of size $10^6$, SP-LSD can perform to a cost which is one order of magnitude lower than the performance of the comparison sorting algorithm. This substancial performance gain is illustrated in the later section of results. 
\newcommand{\tscsort}{
The detailed time complexity some comparison and non-comparison sorting algorithms is given in Table \ref{tab:complexity}. 
\begin{table}[htbp]
\centering
\caption{Complexity analysis of known non-comparison sorting algorithms, where $n$ is input size, $k$ is maximum value, $d$ is number of rounds/digits and $b$ is radix/base for Radix Sort and the number of buckets for Bucket Sort. }
\label{tab:complexity}
\begin{tabular}{|c|c|c|c|c|}
\hline
\textbf{Algorithm} & \textbf{Best Case} & \textbf{Average Case} & \textbf{Worst Case} & \textbf{Space} \\
\hline
Counting Sort & $O(n+k)$ & $O(n+k)$ & $O(n+k)$ & $O(n+k)$ \\
Radix Sort & $O(d \cdot (n+b))$ & $O(d \cdot (n+b))$ & $O(d \cdot (n+b))$ & $O(n+b)$ \\
Bucket Sort & $O(n+b)$ & $O(n+b)$ & $O(n^2)$ & $O(n+b)$ \\
\hline 
\end{tabular}
\end{table}
}
\newcommand{\comparsort}{
We consider the comparison based sorting algorithm, namely, Introsort\LV{ (also called Introspective sort)} \cite{intro-sort97} which is an `in-place'\LV{ (no additional array is required)} but not a `stable'\LV{ (order of appearing repeated elements in the input is not preserved in the output array)} sorting algorithm. 
This is a hybrid sorting algorithm combining quicksort, heapsort and insertion sort. The purpose to combine these algorithms is to tackle the worst-case performance (of quicksort) and to provide faster average case performance. The algorithm begins with quicksort and switches to heapsort when the recursion depth exceeds a level - called {\it depth limit}, based on the logarithm of the number of elements being sorted and switches to insertion sort when the number of elements to be sorted are less. The depth limit is usually chosen as $2*log n$. The algorithm first creates a partition and the size of the partition is compared with the depth limit. If the partition size is greater than a fixed value (many a times, it is 16) and is less than the depth limit, then quicksort is performed. If/When the partition size is greater than the depth limit, then heap sort is performed. If the partition size is too small (than the above said fixed value), then insertion sort is applied. This algorithm uses C++ standard library files to speed-up the average runtime performance. 

There is one more advanced comparison sorting algorithm that is predominantly used for practical applications which is `Timsort'. This is also a hybrid algorithm which uses merge sort and insertion sort used predominantly in Python and Java programs. This is a stable, but not in place algorithm as it requires an additional array and hence uses more space. In this paper, as the sorting algorithms are tested with C++ code, we take Introsort from comparison sorting algorithm (and ignoring Timsort) to benchmark with our sorting algorithm.   
}
\newcommand{\AFS}{
AFS sorting algorithm is a modified version of Radix sort and is designed to sort strings or integers by looking at one byte at a time and assigns it to a group/bucket. The traditional radix sort uses LSD approach and a second `work array' of the same size of the original data to move the items back and forth. On the other hand, AFS performs Most Significant Digit (MSD) approach and swapping concept, it uses no additional memory. AFS partitions the data into buckets. More specifically, AFS groups the item based on their characters and their number of appearances, it forms a `territory'. Then each item is permuted within the original list in the following way. It picks up an item that is in the wrong spot and finds its correct territory. It swaps it with the item that is currently there and then repeats the process with the new item and placing it in the correct place. This forms a `cycle' of swaps that is continued until every item in that section is in its correct place. As AFS avoids duplicating data to an auxiliary array and back, this can run significantly faster on many systems than Radix sort. The issues with AFS are the following: 
\begin{itemize}
\item {\bf Cache Inefficiency:} As established in \cite{LAMARCA99}, bucket-based partitioning induces severe Tanslation Lookaside Buffer (TLB) and cache misses due to random-access memory writes. For simple integers, this latency dominates the runtime.
\item {\bf Fat Bucket Phenomenon:} In skewed distributions, the vast majority of elements cluster into the "zero" bucket. MSD pays the high cost of cache-unfriendly partitioning only to pass nearly the entire dataset to the next recursive level, thus, failing to prune effectively.
\end{itemize}
}

\section{Base-$n$ Radix Sort (BNRS)}
BNRS algorithm is obtained from Radix sort \cite{knuth98} when the base is fixed to $n$ where $n$ is the size of the input. The algorithm in each round, extracts values using $\lfloor A[i] / n^r \rfloor \pmod n$, where $A$ is an array of input values with index $i$ and $r$ is the round number leading to total number of rounds to be
$\lfloor \log_n k \rfloor + 1$, where $k$ is the maximum element of $A$. The motive behind having $n$ as the base is based on the fundamental understanding of the Big $O$ notation which theoretically ignores the constant multiples and makes the calculation strictly relative to the input size $n$. If we maintain the base as $n$, value of $k$ for each round to bound to $n-1$, resulting to, sorting in $O(n)$ time for each round and the total number of rounds is $log_n k$. 
Thus, BNRS achieves an overall time complexity of $O(n \log_n k)$. The weaknesses identified in the algorithm are the following. There is a lack in adaptive control of input in rounds as it performs unnecessary sorting on relatively sorted input amongst the same logarithmic group and the cost for computation overhead for practical performance. We will be addressing them by introducing an optimization approach by partitioning the input array into active region and sorted region with a predicate condition and subsequently reducing the size of the active region.    

Now, we shall explore the novel optimization strategy \SV{that} \LV{utilizing mathematical approach to} effectively reduce the input size for sorting in the consecutive rounds which is achieved by partitioning the input into two regions.
\newcommand{\demonstrationofBNRS}{
In order to understand the BNRS sorting, We demonstrate the sorting with a sample input, where $A = [24, 125, 620, 124, 5]$.
The parameters are: $n=5$ and $k=620$. The number of rounds required is $\lfloor \log_5 620 \rfloor + 1 = 4$.

\paragraph{Round 1 (divisor $n^0=1$)}
\begin{itemize}
    \item Input Array: \texttt{[24, 125, 620, 124, 5]}
    \item Digit Extraction ($\lfloor A[i]/1 \rfloor \pmod 5$):
    \begin{itemize}
        \item[-] $24 \pmod 5 = 4$
        \item[-] $125 \pmod 5 = 0$
        \item[-] $620 \pmod 5 = 0$
        \item[-] $124 \pmod 5 = 4$
        \item[-] $5 \pmod 5 = 0$
    \end{itemize}
    \item Digits for Sorting: \texttt{[4, 0, 0, 4, 0]}
    \item Resultant Array (after applying counting sort): \texttt{[125, 620, 5, 24, 124]}
\end{itemize}

\paragraph{Round 2 (divisor $n^1=5$)}
\begin{itemize}
    \item Input Array: \texttt{[125, 620, 5, 24, 124]}
    \item Digit Extraction ($\lfloor A[i]/5 \rfloor \pmod 5$):
    \begin{itemize}
        \item[-] $\lfloor 125/5 \rfloor \pmod 5 = 25 \pmod 5 = 0$
        \item[-] $\lfloor 620/5 \rfloor \pmod 5 = 124 \pmod 5 = 4$
        \item[-] $\lfloor 5/5 \rfloor \pmod 5 = 1 \pmod 5 = 1$
        \item[-] $\lfloor 24/5 \rfloor \pmod 5 = 4 \pmod 5 = 4$
        \item[-] $\lfloor 124/5 \rfloor \pmod 5 = 24 \pmod 5 = 4$
    \end{itemize}
    \item Digits for Sorting: \texttt{[0, 4, 1, 4, 4]}
    \item Resultant Array: \texttt{[125, 5, 620, 24, 124]}
\end{itemize}
\paragraph{Round 3 (divisor $n^2=25$)}
\begin{itemize}
    \item Input Array: \texttt{[125, 5, 620, 24, 124]}
    \item Digit Extraction ($\lfloor A[i]/25 \rfloor \pmod 5$):
    \begin{itemize}
        \item[-] $\lfloor 125/25 \rfloor \pmod 5 = 5 \pmod 5 = 0$
        \item[-] $\lfloor 5/25 \rfloor \pmod 5 = 0 \pmod 5 = 0$
        \item[-] $\lfloor 620/25 \rfloor \pmod 5 = 24 \pmod 5 = 4$
        \item[-] $\lfloor 24/25 \rfloor \pmod 5 = 0 \pmod 5 = 0$
        \item[-] $\lfloor 124/25 \rfloor \pmod 5 = 4 \pmod 5 = 4$
    \end{itemize}
    \item Digits for Sorting: \texttt{[0, 0, 4, 0, 4]}
    \item Resultant Array: \texttt{[125, 5, 24, 620, 124]}
\end{itemize}

\paragraph{Round 4 (divisor $n^3=125$)}
\begin{itemize}
    \item Input Array: \texttt{[125, 5, 24, 620, 124]}
    \item Digit Extraction ($\lfloor A[i]/125 \rfloor \pmod 5$):
    \begin{itemize}
        \item[-] $\lfloor 125/125 \rfloor \pmod 5 = 1 \pmod 5 = 1$
        \item[-] $\lfloor 5/125 \rfloor \pmod 5 = 0 \pmod 5 = 0$
        \item[-] $\lfloor 24/125 \rfloor \pmod 5 = 0 \pmod 5 = 0$
        \item[-] $\lfloor 620/125 \rfloor \pmod 5 = 4 \pmod 5 = 4$
        \item[-] $\lfloor 124/125 \rfloor \pmod 5 = 0 \pmod 5 = 0$
    \end{itemize}
    \item Digits for Sorting: \texttt{[1, 0, 0, 4, 0]}
    \item Resultant Array: \texttt{[5, 24, 124, 125, 620]}
\end{itemize}

After the final round, the array is fully sorted. The Final Output is \texttt{[5, 24, 124, 125, 620]}.
}
\subsection{Stable Partitioning Approach}
\begin{itemize}
    \item \textbf{Mechanism:} This approach utilizes a stable partitioning algorithm in each round. It separates elements into two groups— `sorted' and `active' based on a  predicate. The predicate evaluates whether an element has been fully sorted relative to the current and future rounds by the condition $A[i] / base^{r-1} = 0$. Elements satisfying this condition are moved to the front, marking distinction between the `sorted' and `active' region. 
    \item \textbf{Advantage:} Only the active portion of the array is processed in the subsequent passes that results in reducing the active portion, though it need not be at every round. This drastically reduces the number of elements handled in later rounds, leading to significant performance gain on skewed datasets 
    \item \textbf{Stability:} The use of a stable partition function and a stable sorting subroutine ensures that the entire algorithm maintains the relative order of equal elements (i.e., the elements having the same values).
\end{itemize}
\LV{\subsection{Sentinel-Based Approach}
\begin{itemize}
    \item \textbf{Mechanism:} This approach involves a one-time, upfront data transformation. Each number is converted into a base-n vector representation of a fixed length. Shorter numbers are padded with a special sentinel value—a digit that cannot normally exist in base-n—to ensure uniform length.
    \item \textbf{Advantage:} Sorted elements are identified by simply checking for the sentinel digit, which replaces complex mathematical calculations in the main sorting loop with a simple equality check. In each round, it builds separate sorted and active lists based on this check.
    \item \textbf{Stability:} The algorithm is fully stable because it appends elements to buckets and reconstructs the array in their original relative order.
\end{itemize}
}
Compared to BNRS, the stable partitioning approach reduces redundant work by limiting the process required for unsorted portion in each round, improving efficiency and performance of the algorithm. However, these improvements come at the cost of increased overhead, which is a drawback, hence we need to find break-even condition which would determine the utility of the optimization in comparison to the cost of implementation by the provision of the early termination strategies.
The notations used in this paper are presented in Table 4, Appendix. 
\section{Analysis of Non-Comparison and Comparison Sorting Algorithms}
In this section, we compare the time complexities of standard comparison-based sorting algorithms \cite{Hoare62} and Base-$n$ Radix Sort (BNRS) to determine boundary of input size from which performance of non-comparison sorting algorithm would outperform comparison sorting algorithms. 
\begin{enumerate}
    \item \textbf{Complexity of Standard Comparison-Based Sorting Algorithms:} Independent of the specific implementation\LV{ (e.g., Merge Sort, Heap Sort, Quick Sort in average case)}, the lower bound for comparison-based sorting algorithms is generally: $T_{\text{comp}} = O(n \log n)$ where $n$ is the number of elements to be sorted and $\log n$ is typically $\log_2 n$. For practical purpose and benchmarking, we use {\it Introsort} \cite{intro-sort97} which is highly efficient, hybrid and stable sorting algorithm designed for practical, real-world data. Details on Introsort is provided in Appendix.  
    \item \textbf{Asymptotic Crossover:} For BNRS (and SP-LSD) to outperform comparison-based sorting algorithms, it's complexity must be asymptotically smaller, satisfying the condition: $O(n \log_n k) < O(n \log n)$\SV{ $\Longrightarrow$} 
    \LV {. Removing the common factor $n$, we get} $\log_n k < \log_2 n$.
    Converting to natural logarithm (ln): $\frac{\ln k}{\ln n} < \frac{\ln n}{\ln 2}$. This gives: 
    $\ln k < \frac{(\ln n)^2}{\ln 2}$. Since, $\frac{(\ln n)}{\ln 2} = log_2 n$, we get $\ln k < \ln n \cdot log_{2} n $. 
    Exponentiating by $e$, $k < e^{(\ln n) \cdot \log_2 n}$.
    Since $e^{\ln n} = n$, we get $k < n^{\log_2 n}$.
    This inequality, $k < n^{\log_2 n}$, defines the condition between $n$ and $k$ under which BNRS and SP-LSD offer a theoretically superior time complexity compared to \LV{standard} comparison-based sorting algorithms.
    
    \item \textbf{Practical analysis of the relation $k < n^{\log_2 n}$:}
    
    We are required to store elements of the dataset in common datatypes. Since we are utilising these datatypes for storing elements into memory, the maximum possible value that can be stored in the datatype becomes our $k_{max}$.
    See Table \ref{tab:crossover} for the crossover input value for different datatypes. A sample mathematical calculation for obtaining the crossover value for the datatype LongLong is discussed in Appendix.   
        
\end{enumerate}
    \newcommand{\datatypecal}{
    \begin{itemize}
        \item Datatype - Int (32-bit): Max value is $2^{31} - 1 \approx 2.147 \times 10^9$
        \item Datatype - Long Long (64-bit): Max value is $2^{63} - 1 \approx 9.223 \times 10^{18}$
    \end{itemize}
    
    Assuming $k_{max}$ to be the maximum possible value for the largest common datatype (64-bit unsigned integer), we can find the crossover point for $n$ as given below:\\
    $k_{max} \le n^{\log_2 n}$. That is, $n^{\log_2 n} \ge 2^{63} - 1$.\\
    Taking $\log_{10}$ of both sides, we get $\log_{10}(n^{\log_2 n}) \ge \log_{10}(2^{63}-1)$. This implies,
    $(\log_2 n) \cdot (\log_{10} n) \ge 18.9648$. Since $\log_2 n = \frac{\log_{10} n}{\log_{10} 2}$, we get     
    $\left(\frac{\log_{10} n}{\log_{10} 2}\right) \cdot (\log_{10} n) \ge 18.9648$. From this we get, $\frac{(\log_{10} n)^2}{0.30103} \ge 18.9648$. That is, $(\log_{10} n)^2 \ge 5.7081 \implies \log_{10} n \ge$ \LV{$\sqrt{5.7081}$} $\approx 2.3891$. This implies,   
    $n \ge 10^{2.3891}$ =   $n \ge 245.004$. Therefore, we get, $n > 245$\\
 This is the theoretical crossover point which implies that when sorting an array containing the maximum possible 64-bit integer value, BNRS algorithm is guaranteed to be theoretically faster than any comparison-based sorting algorithm for average case once the input size surpasses 245 elements.
}
\begin{table}[h!]
\centering
\begin{tabular}{|l|c|c|c|}
\hline
\textbf{Datatype} & \textbf{Bit-Width} & \textbf{Max Value ($k_{max}$)} & \textbf{Crossover ($n >$)} \\
\hline
Int (Signed) & 32-bit & $2^{31} - 1$ & \text{47} \\
Long Long (Signed) & 64-bit & $2^{63} - 1$ & \text{245} \\
BigInt / UUID & 128-bit & $2^{127} - 1$ & \text{2,466} \\
\hline
\end{tabular}
\caption{Practical Crossover values ($n$) for standard datatypes\LV{ derived from $k < n^{\log_2 n}$}.}
\label{tab:crossover}
\end{table}

\section{Optimised Approach – SP-LSD Radix Sort}
Stable Partition - Least Significant Digit Radix Sort or SP-LSD is an optimization algorithm built on the principles of BNRS which simulates the division of the original input array into logical regions, creating a sorted and an active region. The active region of the array is reprocessed at the end of each round, the size of the active region is reduced by tracking the index where all previous elements are considered sorted (\texttt{partition\_ptr}) while leaving the sorted region intact, it contains elements of respective log groups relatively sorted to each other. The decision of whether an element is sorted or not is based on the  condition $(A[i] / base^{r - 1} = 0)$ (refer line 18 of Algorithm 1) satisfied or not, where $r$ is the iteration of rounds, thereby reducing unnecessary the operations on sorted portion of the input. 

In Appendix, we provide the pseudocode of SP-LSD algorithm (Algorithm 1) and Algorithm 2 which is a is a helper function that calls Counting Sort as a subroutine for sorting a portion of the array. We also provide there some details required for understanding Algorithms $1$ and $2$. Further in the Appendix, we demonstrate the SP-LSD algorithm with two examples (one is a bad case example) which will help the readers to understand how SP-LSD works\LV{ in a better way}. 

\newcommand{\Algosandex}{
\begin{algorithm}
\caption{SP-LSD-Sort(A)}
\label{alg:SP-LSD}
\begin{algorithmic}[1]
    \State $n \gets A.length$
    \If{$n \le 1$}
        \State \Return
    \EndIf
    \State $max\_val \gets \textsc{Maximum}(A)$
    \If{$max\_val < 0$}
        \State \Return
    \EndIf
    \State $R \gets \lfloor\log_n(max\_val)\rfloor + 1$
    \If{$R \le 1$}
        \State \textsc{Counting-Sort-by-Digit}(A, A.begin, A.end, 1, n)
        \State \Return
    \EndIf
    
    \State \textsc{Counting-Sort-by-Digit}(A, A.begin, A.end, 1, n)
    \State $first\_active \gets A.begin$
    
    \For{$r \gets 2$ \textbf{to} $R-1$}
        \State $div \gets n^{r-1}$
        \State $partition\_point \gets \textsc{Stable-Partition}(A, first\_active, A.end, \text{where } \lfloor x/div \rfloor = 0)$
        \State \textsc{Counting-Sort-by-Digit}(A, partition\_point, A.end, div, n)
        \State $first\_active \gets partition\_point$
        \If{$first\_active = A.end$}
            \State \Return
        \EndIf
    \EndFor
    
    \State $div \gets n^{R-1}$
    \State \textsc{Counting-Sort-by-Digit}(A, first\_active, A.end, div, n)
\end{algorithmic}
\end{algorithm}

\begin{algorithm}
\caption{Counting-Sort-by-Digit(A, start-iter, end-iter, div, base)}
\label{alg:counting-sort}
\begin{algorithmic}[1]
    \State $n \gets \textsc{Distance}(start-iter, end-iter)$
    \If{$n \le 1$}
        \State \Return
    \EndIf
    \State Let $count[0..base-1]$ be a new array, initialized to 0
    \State Let $output[1..n]$ be a new array
    \For{$it \gets start-iter$ \textbf{to} $end-iter-1$}
        \State $digit \gets \lfloor(\ast it)/div\rfloor \pmod{base}$
        \State $count[digit] \gets count[digit] + 1$
    \EndFor
    \For{$i \gets 1$ \textbf{to} $base-1$}
        \State $count[i] \gets count[i] + count[i-1]$
    \EndFor
    \For{$it \gets end-iter-1$ \textbf{downto} $start-iter$}
        \State $digit \gets \lfloor(\ast it)/div\rfloor \pmod{base}$
        \State $output[count[digit]] \gets \ast it$
        \State $count[digit] \gets count[digit] - 1$
    \EndFor
    \State $i \gets 1$
    \For{$it \gets start-iter$ \textbf{to} $end-iter-1$}
        \State $\ast it \gets output[i]$
        \State $i \gets i + 1$
    \EndFor
\end{algorithmic}
\end{algorithm}
We present some details below for understanding Algorithms $1$ and $2$. 
\begin{itemize}
\item Algorithm $2$ (Counting-Sort-by-Digit) is a helper function that calls Counting Sort as a subroutine for sorting a portion of the array, here called for the active region.

\item Algorithm $1$ (SP-LSD-Sort) breaks down the sorting process into $3$ distinct stages determined by the rounds, total number of rounds is calculated by $R \gets \lfloor\log_n(max\_val)\rfloor + 1$ (see line 9 of Algorithm 1). The first stage applies sorting to the entire input, the second stage runs the core optimization  by partitioning the array into 'sorted' and 'active' regions using $first\_active$ iterator to mark the active region, $div \gets n^{r-1}$ (see line 17 of Algorithm 1) where $r$ is iterated value from $2$ to $R-1$, Stable-Partition is called from $first\_active$ to the 'past-the-end' iterator. The final stage is the round $R-1$, where Counting Sort is called for the remaining active region.
\end{itemize}
We now present two examples where the second example is a bad case for a poor input.  
\subsection{Example1: Demonstration of SP-LSD}
\textbf{Input:} $A = [4, 1, 620, 124, 3]$, Base ($n$): 5, Max Value ($k$): 620, Rounds Needed: $\lfloor\log_5(620)\rfloor + 1 = 4$.

\paragraph{Initial State}
The array is in its original unsorted state. \texttt{first\_active\_idx = 0}.
\begin{center}
\texttt{\begin{tabular}{|c|c|c|c|c|}
\hline
0 & 1 & 2 & 3 & 4 \\
\hline
4 & 1 & 620 & 124 & 3 \\
\hline
\end{tabular}}
\end{center}

\paragraph{Round 1 (div=1)}
The first round sorts by the least significant digit ($\text{val} \pmod 5$). The active range is \texttt{A[0,4]}.
\begin{itemize}
    \item {Digits:} \texttt{[4, 1, 0, 4, 3]}
    \item {Result (after sorting):} The \texttt{first\_active\_idx} remains \texttt{0}.
\end{itemize}
\begin{center}
\texttt{\begin{tabular}{|c|c|c|c|c|}
\hline
0 & 1 & 2 & 3 & 4 \\
\hline
620 & 1 & 3 & 4 & 124 \\
\hline
\end{tabular}}
\end{center}

\paragraph{Round 2 (div=5)}
The active range is \texttt{A[0,4]}. The partition predicate is $\lfloor x / 5 \rfloor = 0$. This is TRUE for \texttt{1}, \texttt{3}, and \texttt{4}.
\newline
{State after Partitioning:} \texttt{STABLE-PARTITION} moves the sorted elements to the 'sorted' region, partition is set at index 3.
\begin{center}
\texttt{\begin{tabular}{|c|c|c|c|c|}
\hline
0 & 1 & 2 & 3 & 4 \\
\hline
1 & 3 & 4 & 620 & 124 \\
\hline
\multicolumn{3}{|c|}{Sorted} & \multicolumn{2}{c|}{Active} \\
\hline
\end{tabular}}
\end{center}
{Action:} \texttt{COUNTING-SORT} now runs on the active range \texttt{A[3,4]}.
\begin{itemize}
    \item {Digits ($\lfloor \text{val}/5 \rfloor \pmod 5$):} \texttt{[4, 4]}
    \item {Result:} The active range is sorted.
\end{itemize}
\begin{center}
\texttt{\begin{tabular}{|c|c|c|c|c|}
\hline
0 & 1 & 2 & 3 & 4 \\
\hline
1 & 3 & 4 & 620 & 124 \\
\hline
\end{tabular}}
\end{center}

\paragraph{Round 3 (div=25)}
Active Range: \texttt{A[3,4]}. The predicate is $\lfloor x / 25 \rfloor = 0$. This is FALSE for both \texttt{620} and \texttt{124}.
\newline
{Action:} \texttt{COUNTING-SORT} runs on \texttt{A[3,4]}.
\begin{itemize}
    \item {Digits ($\lfloor \text{val}/25 \rfloor \pmod 5$):} \texttt{[4, 4]}
    \item {Result:} The active range is sorted. 
\end{itemize}
\begin{center}
\texttt{\begin{tabular}{|c|c|c|c|c|}
\hline
0 & 1 & 2 & 3 & 4 \\
\hline
1 & 3 & 4 & 620 & 124 \\
\hline
\end{tabular}}
\end{center}

\paragraph{Round 4 (div=125)}
The active range is still \texttt{A[3,4]}. The predicate is $\lfloor x / 125 \rfloor = 0$. This is TRUE for \texttt{124}.
\newline
{State after Partitioning:} The partition moves \texttt{124} to the 'sorted' region. 
\begin{center}
\texttt{\begin{tabular}{|c|c|c|c|c|}
\hline
0 & 1 & 2 & 3 & 4 \\
\hline
1 & 3 & 4 & 124 & 620 \\
\hline
\multicolumn{4}{|c|}{Sorted} & \multicolumn{1}{c|}{Active} \\
\hline
\end{tabular}}
\end{center}
{Action:} \texttt{COUNTING-SORT} runs the final round on index \texttt{A[4,4]}, the algorithm terminates after completion of this stage.

\paragraph{Final Sorted Array}
\begin{center}
\texttt{\begin{tabular}{|c|c|c|c|c|}
\hline
0 & 1 & 2 & 3 & 4 \\
\hline
1 & 3 & 4 & 124 & 620 \\
\hline
\end{tabular}}
\end{center}

We now demonstrate how SP-LSD sorts with the same sample input used to demonstrate BNRS, this demonstration is a poor example, reasoning is discussed further along in the note section. \\

\textbf{Input:} $A = [24, 125, 620, 124, 5]$, Base ($n$): 5, Max Value ($k$): 620, Rounds Needed: $\lfloor\log_5(620)\rfloor + 1 = 4$.

\paragraph{Initial State}
The \texttt{first\_active\_idx} marks the start of the subarray to be processed. Initially, this is the entire array. \texttt{first\_active\_idx = 0}.
\begin{center}
\texttt{\begin{tabular}{|c|c|c|c|c|}
\hline
0 & 1 & 2 & 3 & 4 \\
\hline
24 & 125 & 620 & 124 & 5 \\
\hline
\end{tabular}}
\end{center}

\paragraph{Round 1 (div=1)}
Active Range: \texttt{A[0,4]}. 
The array is unchanged. The new partition starts at index 0. Action: \texttt{COUNTING-SORT} on the active range \texttt{A[0,4]}. Digits ($\text{val} \pmod 5$): \texttt{[4, 0, 0, 4, 0]}.
\newline
Result: \texttt{first\_active\_idx = 0} (No elements were finalized).
\begin{center}
\texttt{\begin{tabular}{|c|c|c|c|c|}
\hline
0 & 1 & 2 & 3 & 4 \\
\hline
125 & 620 & 5 & 24 & 124 \\
\hline
\end{tabular}}
\end{center}

\paragraph{Round 2 (div=5)}
Active Range: \texttt{A[0,4]}. Partitioning: \texttt{STABLE-PARTITION} is called, the predicate $\lfloor x / 5 \rfloor = 0$ is false for all elements. The array is unchanged. The new partition starts at index 0. Action: \texttt{COUNTING-SORT} on the active range \texttt{A[0,4]}. Digits ($\lfloor \text{val}/5 \rfloor \pmod 5$): \texttt{[0, 4, 1, 4, 4]}.
\newline
Result: \texttt{first\_active\_idx = 0} (No elements were finalized).
\begin{center}
\texttt{\begin{tabular}{|c|c|c|c|c|}
\hline
0 & 1 & 2 & 3 & 4 \\
\hline
125 & 5 & 620 & 24 & 124 \\
\hline
\end{tabular}}
\end{center}

Note: For the break-even condition we needed the input to have atleast half of the elements to be partitioned in second round. This causes the cost of optimization to exceed the benefit provided in terms of number of operations.

\paragraph{Round 3 (div=25)}
Active Range: \texttt{A[0,4]}. Partitioning: \texttt{STABLE-PARTITION} is called. The predicate is $\lfloor x / 25 \rfloor = 0$. This is TRUE for \texttt{5} and \texttt{24}. The partition preserves the original relative order. The array becomes \texttt{[5, 24, 125, 620, 124]}. The new partition separating finalized and active elements is at index 2.
\newline
State before Sorting:
\begin{center}
\texttt{\begin{tabular}{|c|c|c|c|c|}
\hline
0 & 1 & 2 & 3 & 4 \\
\hline
5 & 24 & 125 & 620 & 124 \\
\hline
\multicolumn{2}{|c|}{Sorted} & \multicolumn{3}{c|}{Active} \\
\hline
\end{tabular}}
\end{center}
Action: \texttt{COUNTING-SORT} on the new active range \texttt{A[2,4]}. Digits ($\lfloor \text{val}/25 \rfloor \pmod 5$): \texttt{[0, 4, 4]}. The sorted active range is: \texttt{[125, 620, 124]}.
\newline
Result: \texttt{first\_active\_idx = 2} (Updated for the next round).
\begin{center}
\texttt{\begin{tabular}{|c|c|c|c|c|}
\hline
0 & 1 & 2 & 3 & 4 \\
\hline
5 & 24 & 125 & 620 & 124 \\
\hline
\end{tabular}}
\end{center}

\paragraph{Round 4 (div=125)}
Active Range: \texttt{A[2,4]}. Partitioning: \texttt{STABLE-PARTITION} is called on this sub-array. Predicate is $\lfloor x / 125 \rfloor = 0$. This is TRUE for \texttt{124}. The sub-array is partitioned to \texttt{[124, 125, 620]}. The full array \texttt{A} is now \texttt{[5, 24, 124, 125, 620]}. The new partition is at index 3.
\newline
State before Sorting:
\begin{center}
\texttt{\begin{tabular}{|c|c|c|c|c|}
\hline
0 & 1 & 2 & 3 & 4 \\
\hline
5 & 24 & 124 & 125 & 620 \\
\hline
\multicolumn{3}{|c|}{Sorted} & \multicolumn{2}{c|}{Active} \\
\hline
\end{tabular}}
\end{center}
Action: \texttt{COUNTING-SORT} on the new active range \texttt{A[3,4]}. Digits ($\lfloor \text{val}/125 \rfloor \pmod 5$): \texttt{[1, 4]}. The sorted active range is \texttt{[125, 620]}.
\newline
Result: \texttt{first\_active\_idx = 3}.
\begin{center}
\texttt{\begin{tabular}{|c|c|c|c|c|}
\hline
0 & 1 & 2 & 3 & 4 \\
\hline
5 & 24 & 124 & 125 & 620 \\
\hline
\end{tabular}}
\end{center}

\paragraph{Final Sorted Array}
After the final round, the array is fully sorted.
\begin{center}
\texttt{\begin{tabular}{|c|c|c|c|c|}
\hline
0 & 1 & 2 & 3 & 4 \\
\hline
5 & 24 & 124 & 125 & 620 \\
\hline
\end{tabular}}
\end{center}
}
\section{Analysis of the Optimization}
We can break down the cost of operations of the algorithm in stages on the basis of rounds, which run from 1 to $R$. The number of rounds is determined by the formula $\lfloor \log_n k \rfloor + 1$, where $k$ is the maximum element. We further define $\alpha$ to be the cost for sorting and $\beta$ to be the cost for the partition and $c = \alpha / \beta $. Also, we define $p$ as the proportion of elements partitioned-out after  the first effective partitioning step. Then,   
\begin{itemize}
    \item {Stage 1} (Round 1): Perform sorting on the mod values of the entire array, no partitioning required.
    \item {Stage 2 }(Rounds 2 to $R-1$): Perform partitioning and sort the active region iteratively.
    \item Stage 3 (Round $R$): Perform sorting on the final active region, no partitioning required.
\end{itemize}
\subsection{The Pruning Crossover}
Introducing adaptiveness provides a significant trade-off: SP-LSD incurs structural overheads (partitioning cost) to prune redundant operations (sorting cost). Unlike the deterministic BNRS, SP-LSD's cost is inherently input-dependent, governed by the shrinkage rate of the active region. To theoretically bound this behavior, we perform a rigorous cost-benefit analysis, distinguishing between fixed full dataset operations and varying active region operations through the defined stages. We evaluate this trade-off under two boundary conditions defined by dataset assumptions:

\begin{enumerate}
    \item \textbf{Skewed Distribution:} Analysis of the Skewed Distribution provides the minimum required skew ($p$) for the viability of performing SP-LSD.
    \item \textbf{Uniform Log Distribution:} Analysis of the Uniform Log Distribution defines the hardware cost limits ($c$) required for the algorithm to win on uniform partitioning.
\end{enumerate}
\subsection{Theoretical Analysis and Performance Prediction}

 Modelling the theoretical operational cost on a large input with skewed data:
\paragraph{Parameters:}
\begin{itemize}
    \item $n$ (input size): $10^6$ (1,000,000)
    \item $k$ (maximum value): $9 \times 10^{18}$ (the approximate max value for a 64-bit integer)
    \item Assuming $1\%$ distribution of elements among each log group after the first log group.
\end{itemize}

For these parameters, the number of rounds required is $R=4$, calculated from $\lfloor\log_{10^6}(10^{18})\rfloor + 1 = 4$. We assume a skewed dataset where maximum of 97\% of the elements are lesser than $n$, giving a partition proportion $p = 0.97$, which is the best case for our partitioning logic (assuming 1\% of input size in each log distribution).

\paragraph{Cost of Comparison-Based Sort (\texttt{Introsort}):}
The lower bound for comparison-based sorting is $T_{\text{comp}}=\Omega(n \log n)$, calculating on which gives us the following number of operations: 
\[ T_{\text{comp}} \approx n \log_2 n = 10^6 \times \log_2(10^6) \approx 10^6 \times 19.93 = {19,930,000~{\text{operations}}}. \]

\paragraph{Cost of BNRS:}
The cost is fixed, as the algorithm processes all $n$ elements in each of $R$ rounds.
\[ T_{\text{BNRS}} = R \times n = 4 \times 10^6 = 4 \times 1,000,000 = {4,000,000~{\text{operations}}}. \]

\paragraph{Cost of SP-LSD:} Total cost of operation assuming skewed distribution case
\begin{align*}
    T_{\text{SP-LSD}} &= \underbrace{1,000,000}_{\text{Round 1 (Sort)}} + \underbrace{(1,030,000 + 60,000)}_{\text{Rounds 2-3 (Partition+ Sort)}} + \underbrace{30,000}_{\text{Round 4 (Sort)}} \\
    &= {2,120,000~\text{operations}}.
\end{align*}

This analysis demonstrates that the BNRS to be theoretically five times faster than an optimal comparison-based sort and the SP-LSD optimization then further provides a two times improvement over BNRS. The cumulative effect is a predicted ten times performance advantage for SP-LSD over standard comparison-based sorting.
\subsection{Round Feasibility Crossover}
The efficacy of SP-LSD is not uniform across execution; it is architecturally confined to the intermediate rounds. By analyzing the execution stages and theoretical modelling, we observe that:
\begin{itemize}
    \item \textbf{Stage 1 (Initialization):} Incurs a fixed cost ($\alpha N$) identical to BNRS. No pruning is possible here as the full dataset must be read to establish the first digit's order.
    \item \textbf{Stage 3 (Termination):} Incurs a variable cost but offers no further savings, as there is no subsequent round to benefit from a partition.
    \item \textbf{Stage 2 (The Core):} This is the exclusive window for optimization. The cost of partitioning in round $r$ is amortized only if it reduces the sorting workload in round $r+1$.
\end{itemize}

Consequently, if the total number of rounds $R \le 2$, the "Optimization Window" (Stage 2) is effectively empty. In such cases, SP-LSD incurs the overhead of Stage 1 and Stage 3 without any intermediate rounds to realize the gain from partitioning. Therefore, a structural prerequisite for SP-LSD viability is the existence of at least one intermediate round, implying $R > 2$, from this we arrive to our crossover, namely \textbf{Round-feasibility Crossover} ($k > n^2$). Accordingly, our analytical results and viability tables presented in the subsequent sections begin evaluation at $R=3$.

\subsection{Case 1: Conservative Skew Model}
We first derive the condition for datasets exhibiting heavy tail skew. We adopt a conservative assumption where effective pruning occurs in the first partition (Round 2), reducing the active set to $(1-p)n$. For the remainder of Stage 2 and Stage 3, we conservatively assume the active region does not shrink further.
\vskip .2cm
\textbf{Cost Derivation:}
\begin{itemize}
    \item \textbf{Stage 1 (Round 1):} Perform sorting on the mod values of the entire array, no partitioning required.
     $\text{Cost}_{\text{Stage 1}} = \alpha \cdot n $
    
    \item \textbf{Stage 2 (Rounds $2 \dots R-1$):} Perform partitioning and sort the active region iteratively. Note that the partition in Round 2 acts on the full array size $n$ to identify zeros, while subsequent operations act on the reduced active set $(1-p)n$.
    \begin{itemize}
        \item Round 2 Cost: $\beta \cdot n$ (full partition) + $\alpha(1-p)n$ (partial sort).
        \item Rounds $3 \dots R-1$ Cost: Total of $R-3$ rounds (excluding the first two rounds and the last round $R$) involving both partitioning and sorting on the active set. Cost: $(R-3)(\alpha + \beta)(1-p)n.$
    \end{itemize}
    $$ \text{Cost}_{\text{Stage 2}} = \beta n + \alpha(1-p)n + (R-3)(\alpha + \beta)(1-p)n. $$

    \item \textbf{Stage 3 (Round $R$):} Perform sorting on the final active region, no partitioning required.
     $\text{Cost}_{\text{Stage 3}} = \alpha(1-p)n. $
\end{itemize}

\textbf{Total Cost Equation:}
Summing Stage 1 with the combined costs of Stage 2 and 3:
$T_{\text{SP-LSD}}^{\text{skewed}} = n[ (\alpha + \beta)  + \left[ (R-1)\alpha + (R-3)\beta \right] (1-p) ]. $

\textbf{Break-Even Analysis:}
We compare this against standard BNRS ($T_{\text{BNRS}} = R \cdot \alpha \cdot n$). Substituting $\alpha = c\beta$, and cancelling the common terms $n$ and $\beta$\SV{:}\LV{ on both sides, we get}
\begin{align*}
(c+1) + [ c(R-1) + (R - 3) ] (1-p) &< Rc \\
[ c(R-1) + (R - 3) ] (1-p) &< Rc - c - 1 \\
1-p &< \frac{c(R-1) - 1}{c(R-1) + (R - 3)}
\end{align*}
Solving for $p$, we obtain the \textbf{Skew Crossover}:
\begin{equation*}
\label{eq:skew_crossover}
p > \frac{R-2}{c(R-1) + (R - 3)}
\end{equation*}

\begin{table}[h]
    \centering
    \caption{Minimum Pruning Fraction ($p$) Required for SP-LSD Viability across varying Rounds ($R$) and Cost Ratios ($c$). Calculated using the Pruning Crossover inequality $p > \frac{R-2}{c(R-1) + R - 3}$.}
    \label{tab:pruning_crossover_values}
    \vspace{0.2cm}
    \begin{tabular}{cccccc}
        \toprule
        \textbf{R (Rounds)} & \multicolumn{5}{c}{\textbf{Cost Ratio} ($c = \alpha/\beta$)} \\
        \cmidrule(lr){2-6}
         & $\mathbf{c=1}$ & $\mathbf{c=2}$ & $\mathbf{c=3}$ & $\mathbf{c=4}$ & $\mathbf{c=5}$ \\
        \midrule
        3  & 0.500 & 0.250 & 0.167 & 0.125 & 0.100 \\
        4  & 0.500 & 0.286 & 0.200 & 0.154 & 0.125 \\
        8  & 0.500 & 0.316 & 0.231 & 0.182 & 0.150 \\
        16 & 0.500 & 0.326 & 0.241 & 0.192 & 0.159 \\
        32 & 0.500 & 0.330 & 0.246 & 0.197 & 0.163 \\
        \bottomrule
    \end{tabular}
\end{table}
\subsection{Case 2: Uniform Log Distribution} 
We analyze performance on a uniform input distribution within log groups, where the active set shrinks by a constant fraction $1/R$ in each round. We assume the active set size at round $r$ is $n_r = n(1 - \frac{r-1}{R})$.

\noindent \textbf{Cost Derivation:}
\begin{itemize}
    \item \textbf{Stage 1:} Full Sort. Cost $= \alpha \cdot n$.
    \item \textbf{Stage 2 (Iterative Sum):} Rounds $2$ to $R-1$ involve both partitioning and sorting on the decaying active set $n_r$.
    $$ \text{Cost}_{\text{Stage 2}} = (\alpha + \beta) \sum_{r=2}^{R-1} n \left(1 - \frac{r-1}{R}\right) $$
    Using the discrete sum identity for the range $r=2 \dots R-1$:
    $$ \text{Cost}_{\text{Stage 2}} = (\alpha + \beta) n \left[ (R-2) - \frac{(R-2)(R-1)}{2R} \right] $$
    \item \textbf{Stage 3:} Sort remaining elements $n/R$. Cost $=\alpha (n/R)$.
\end{itemize}
\textbf{Total Cost Equation:}
Summing Stage 1 with the combined costs of Stage 2 and 3, we get, 
$T_{\text{SP-LSD}}^{\text{uniform}} = \alpha n + (\alpha + \beta) n \left[ (R-2) - \frac{(R-2)(R-1)}{2R} \right] + \alpha (n/R).$ 
\vskip .2cm 
\textbf{Break-Even Analysis:}
We require $T_{\text{SP-LSD}}^{\text{uniform}} < T_{\text{BNRS}}$. Following the summation and simplification (note $T_{\text{BNRS}} = R \cdot \alpha \cdot  n$):
$$ (\alpha + \beta) \left[ \frac{R^2 - R - 2}{2R} \right] < \alpha \left( \frac{R^2 - R - 1}{R} \right) $$
Substituting $\alpha = c\beta$ and solving for $c$:
\begin{equation}
c > \frac{R^2 - R - 2}{R^2 - R} = 1 - \frac{2}{R(R-1)}
\end{equation}

Solving for $R$, 
\begin{equation}
    R < R_{\max}(c) = \frac{1 + \sqrt{1 + \frac{8}{1-c}}}{2}
\end{equation}

\hspace*{-1cm}

\begin{table}[h]
\centering

\begin{minipage}{0.48\textwidth}
\centering
\scalebox{0.8}{
\begin{tabular}{|c|c|}
    \hline
    \textbf{} & \textbf{Min. Cost Ratio} \\
    \textbf{Rounds R} & ($c > \dots$) \\
    \hline
    3  & {0.667} \\
    4  & {0.833} \\
    8  & {0.964} \\
    16 & {0.992} \\
    32 & {0.998} \\
    \hline
\end{tabular}}
\caption{Minimum Cost Ratio ($c$) Required for SP-LSD Viability.}
\end{minipage}
\hfill
\begin{minipage}{0.48\textwidth}
\centering
\scalebox{0.8}{
\begin{tabular}{|c|c|c|}
    \hline
    \textbf{Cost Ratio} & \textbf{Upper Bound} & \textbf{Maximum} \\
    ($c = \alpha/\beta$) & ($R < \dots$) & \textbf{Integer Value ($k$)} \\
    \hline
    0.95 & 6.85 & $n^6-1$ \\
    0.90 & 5.00 & $n^4-1$ \\
    0.85 & 4.19 & $n^4-1$ \\
    0.80 & 3.70 & $n^3-1$ \\
    0.75 & 3.37 & $n^3-1$ \\
    0.70 & 3.13 & $n^3-1$ \\
    0.67 & 3.01 & $n^3-1$ \\
    \hline
\end{tabular}}
\caption{Maximum Feasible Rounds ($R_{\max}$) for Suboptimal Hardware ($c < 1$).}
\end{minipage}

\end{table}

\subsection{Observations:}
\begin{itemize}
\item Table 3 establishes the minimum hardware cost ratio required for viability as the number of rounds increases, confirming effectiveness for all standard integer widths where $R>2$. 

\item Table 4 demonstrates the algorithm's robustness against poor hardware efficiency, defining the maximum sortable magnitude ($n^R$) sustainable when partitioning is more expensive than sorting.
\end{itemize}

\section{Experimental Evaluation and Results}

To validate the efficiency of SP-LSD, we conducted both formal theoretical operational cost analysis and experimental benchmark results. We compare the performance of the SP-LSD against Base-$n$ Radix Sort (BNRS), standard Base-10 Radix Sort (Baseline Radix Sort) and optimized comparison-based sort from the C++ standard library, \texttt{std::sort}.

\subsection{Experimental Setup}

All benchmarks were conducted on a consistent hardware and software environment to ensure the comparability of results measured using Google Benchmark Library. The specifications of the test machine are detailed in Table~\ref{tab:system_specs}.

\begin{table}[htbp]
\centering
\caption{System specifications for benchmarking environment.}
\label{tab:system_specs}
\begin{tabular}{@{}ll@{}}
\toprule
\textbf{Component} & \textbf{Specification} \\
\midrule
Processor & 11th Gen Intel(R) Core(TM) i5-1135G7 @ 2.40GHz \\
Memory (RAM) & 12.0 GB \\
Operating System & Ubuntu 22.04 LTS (64-bit) \\
Compiler & GCC 11.4 with -O3 flag \\
\bottomrule
\end{tabular}
\end{table}

\subsection{Empirical Benchmarks on Skewed Data}

The results presented in Table~\ref{tab:benchmark_results} were generated using a skewed dataset of 64-bit integers, representing the ideal use-case for our optimization.

\noindent To rigorously evaluate SP-LSD against realistic high-performance workloads without relying on artificial random skew, we utilized a `System-Calibrated' benchmark modeling \textbf{Heavy-Tailed Delta-Encoded Streams} (e.g., compressed logs). Furthermore, to strictly enforce the structural requirement of supporting the full 64-bit range ($k > n^2$) and prevent naive global-clipping optimizations, we explicitly injected 1\% outliers of the maximum 64-bit integer ($2^{64}-1$). This configuration generates a realistic bitwise density profile. Additionally, to ensure benchmark reflects optimal performance, we apply bitwise operations to both BNRS and SP-LSD implementation rather than computationally expensive modulo and division operations. This allows for a direct validation of the SP-LSD algorithm's adaptive partitioning strategy. 

\begin{figure}[htbp]
\centering
\begin{subfigure}{0.35\textwidth}
\includegraphics[width=6cm,height=4cm]{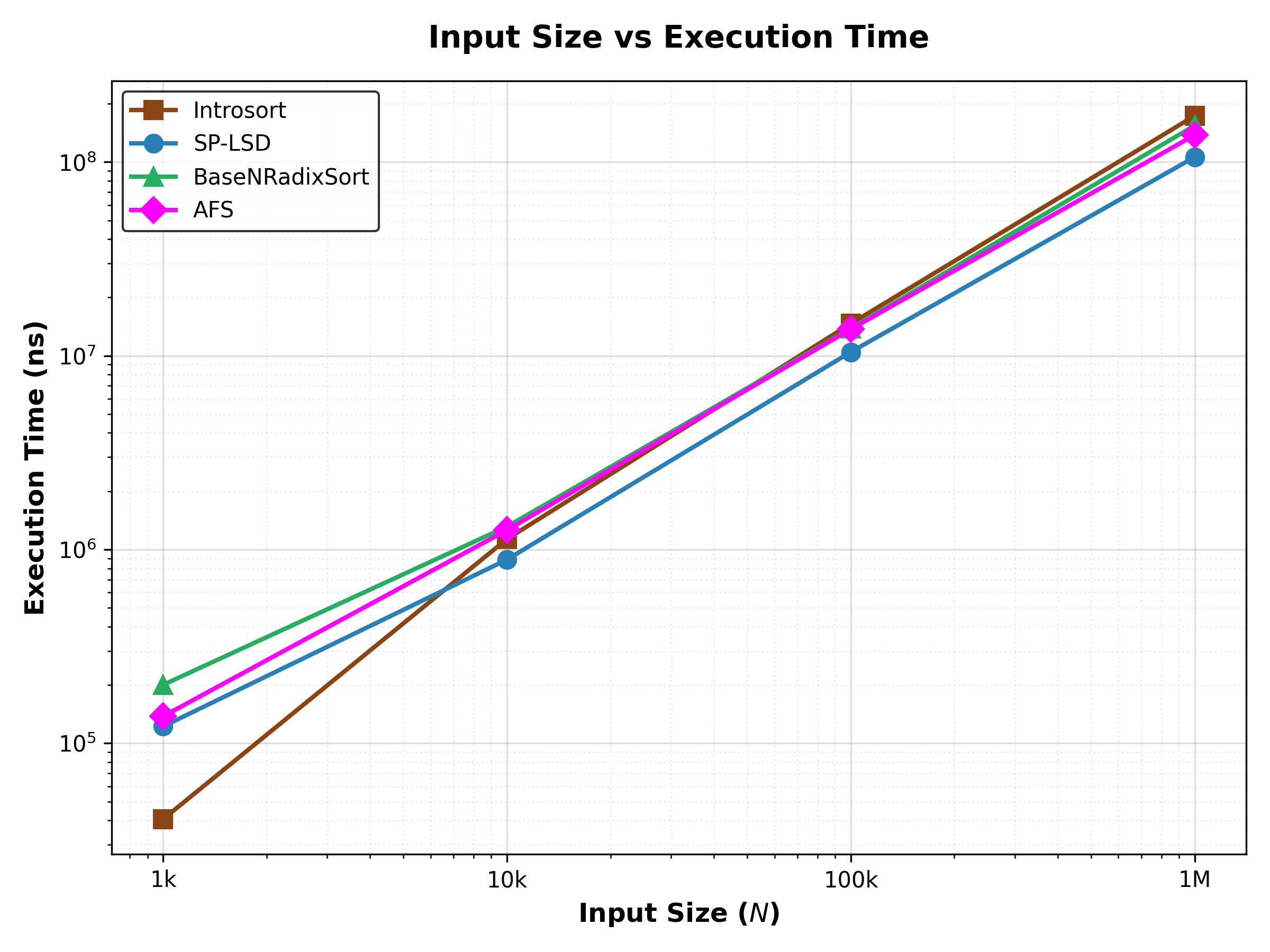}
\caption{Performance of sorting algorithms: Input size vs Execution Time.}
\label{fig-input-vs-time}
\end{subfigure}
\hspace*{1cm}
\begin{subfigure}{0.45\textwidth}
\includegraphics[width=6.5cm,height=5cm]{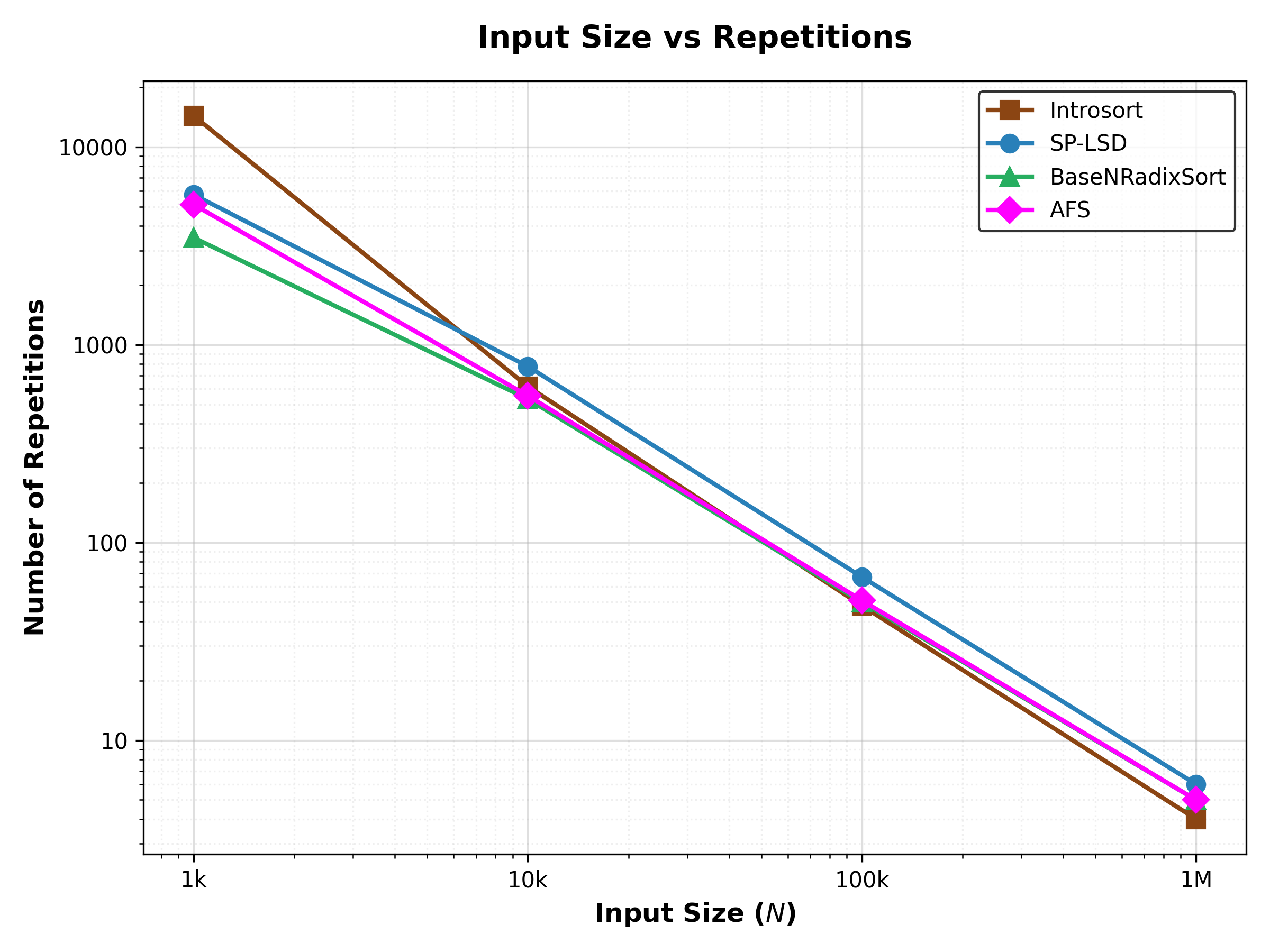}
\caption{Input size and repetition of sorting in a fixed time.} 
\label{fig-repeat}
\end{subfigure}
\caption{ }
\end{figure}

\begin{figure}[htbp]
\centering
\begin{subfigure}{0.40\textwidth}
\includegraphics[width=6cm,height=5cm]{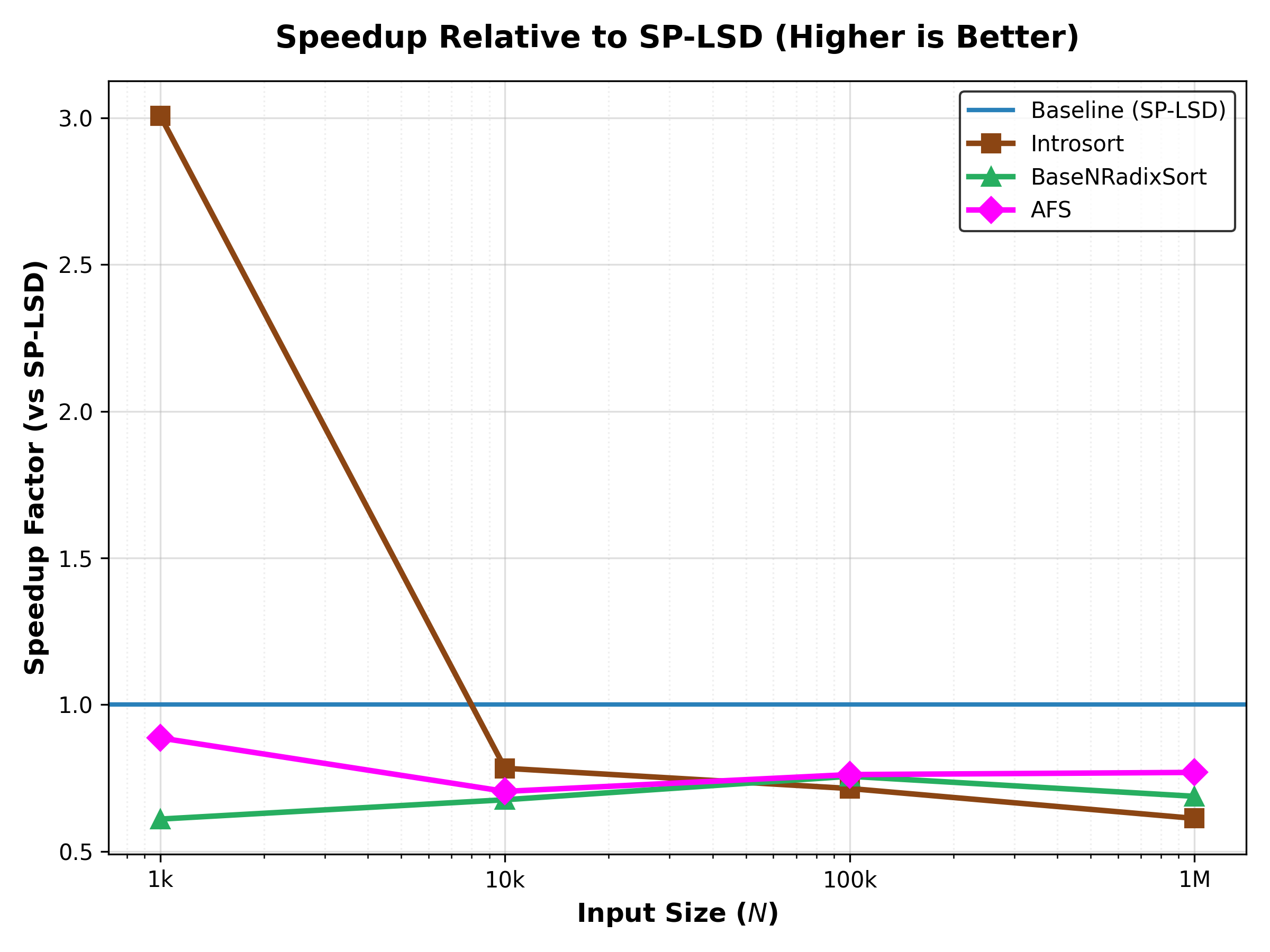}
\caption{Speed-up of other algorithms with SP-LSD}
\label{fig-speed-up}
\end{subfigure}
\hspace*{1cm}
\begin{subfigure}{0.40\textwidth}
\includegraphics[width=6.5cm,height=5cm]{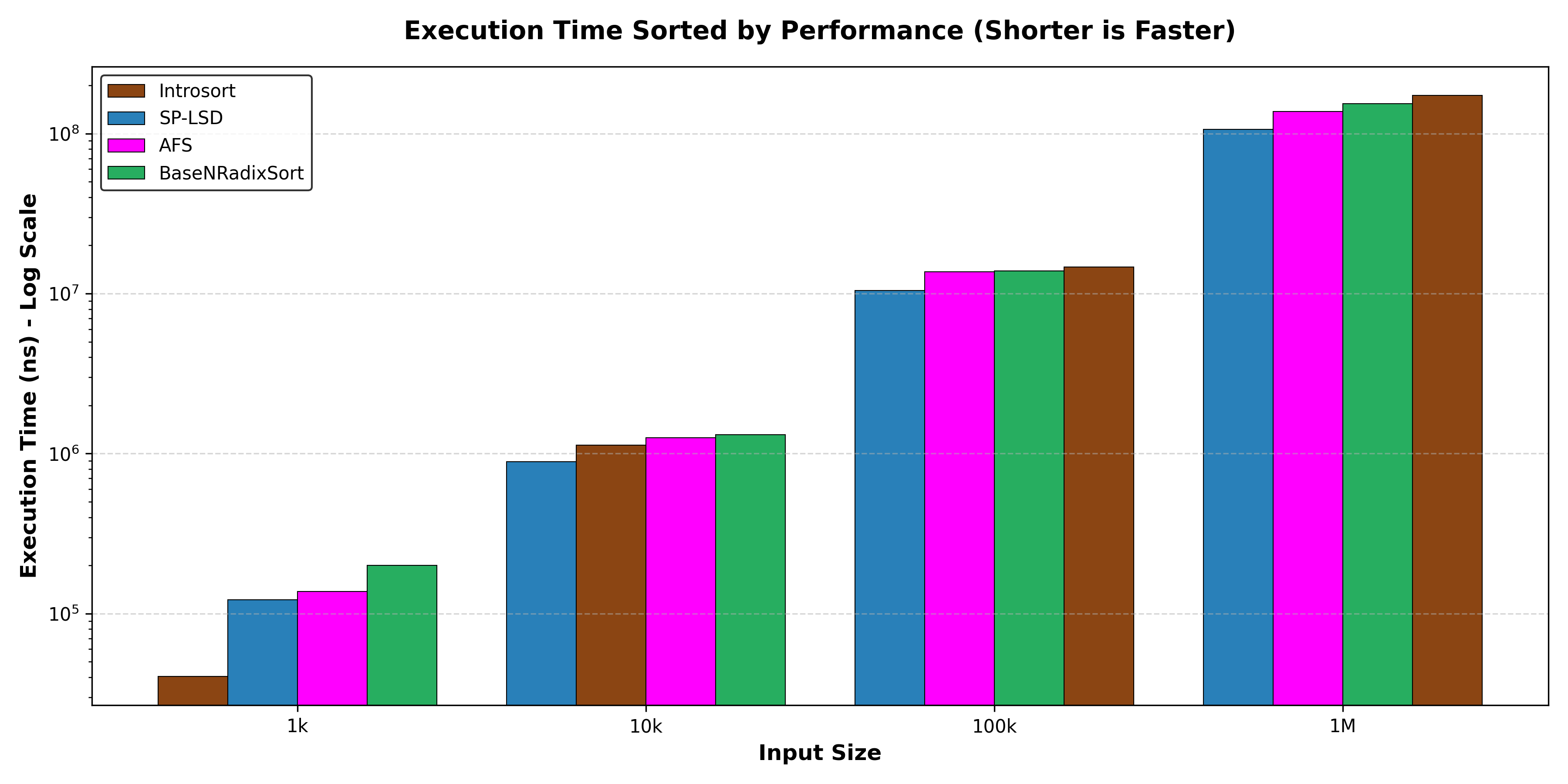}
\caption{Execution time of sorting algorithms ordered by performance-wise} 
\label{fig-performance-sort}
\end{subfigure}
\caption{ } 
\end{figure}

\begin{table}[h]
\centering
\scalebox{0.8}{
\begin{tabular}{|l|r|r|r|r|c|}
\hline
\textbf{Parameter} & \textbf{Introsort} & \textbf{AFS} & \textbf{BNRS} & \textbf{SP-LSD} & \textbf{Input Size} \\
\hline
Average Time (ns) & 40,657 & 137,911 & 200,452 & 122,265 & \multirow{3}{*}{1,000} \\
CPU Time (ns)     & 40,647 & 137,908 & 200,423 & 122,253 & \\
Repetitions       & 14,381 & 5,116   & 3,480   & 5,746   & \\
\hline
Average Time (ns) & 1,133,385 & 1,260,366 & 1,313,428 & 887,678 & \multirow{3}{*}{10,000} \\
CPU Time (ns)     & 1,132,968 & 1,260,277 & 1,313,279 & 887,553 & \\
Repetitions       & 616       & 553       & 532       & 778       & \\
\hline
Average Time (ns) & 14,636,518 & 13,725,100 & 13,829,398 & 10,452,841 & \multirow{3}{*}{100,000} \\
CPU Time (ns)     & 14,628,757 & 13,722,726 & 13,826,997 & 10,450,737 & \\
Repetitions       & 48         & 51         & 50         & 67         & \\
\hline
Average Time (ns) & 173,054,556 & 137,778,640 & 154,112,030 & 105,962,703 & \multirow{3}{*}{1,000,000} \\
CPU Time (ns)     & 173,046,416 & 137,762,253 & 153,749,641 & 105,946,853 & \\
Repetitions       & 4           & 5           & 5           & 6           & \\
\hline
\end{tabular}}
\caption{Benchmark results for Introsort, American Flag Sort (AFS), Base-$n$ Radix Sort (BNRS), and SP-LSD on skewed datasets.}
\label{tab:benchmark_results}
\end{table}

\subsection{Analysis of Results}
The benchmark results from the skewed dataset consisting of metrics like input size, execution time, speed-up factor, repetitions, are provided in Fig. \ref{fig-input-vs-time}, Fig. \ref{fig-repeat}, Fig. \ref{fig-speed-up} and Fig. \ref{fig-performance-sort}. the metric repetitions refers to given a time bound, how many times the whole algorithm is executed repeatedly. Table \ref{tab:benchmark_results} refers to the results of executing all the sorting algorithms with respect to the parameters average time, CPU time and repetitions. consisting the parameters. The average time is calculated based on the sorting algorithm tested for five different random inputs generated on the basis of random input generated by the dataset.
The results can be seen or accessed in: 
\url{https://github.com/Atharv-Pandey/SP-LSD_Benchmark}. With the obtained values from the above said figures and tables, the core findings are as follows:
\begin{enumerate}
    \item \textbf{Performance at Small Scales:}
    At small input sizes ($n=1,000$), Introsort (40.7 $\mu$s) reigns supreme, significantly outperforming all radix-based variants ($>120 \mu$s). This validates the theoretical initial startup costs associated with non-comparison sorting (buffer allocation, frequency counting) which only amortize at larger scales. This observation establishes the lower bound of the crossover region where simpler comparison sorts are preferable. However, we also notice that despite theoretical proof suggesting a better performance derived from ($k < n^{\log_2 n}$) condition resulting in $n > 245$ for $64$ bit integer, performance gains were realised only in the input range exceeding $10,000$. This delay is caused due to initialization of overhead cost while implementing non-comparison sorting algorithms.

    \item \textbf{Large-Scale Supremacy of Non-comparison Sorting:}
    For large datasets ($n=1,000,000$), BNRS (154.1 ms) establishes a clear lead over Introsort (173.1 ms). Notably, at $N=100,000$, the performance was comparable (13.8 ms vs 14.6 ms), but the gap widens at $10^6$, confirming the asymptotic advantage of the non-comparison approach $O(n)$ over the comparison-based one $O(n \log n)$ as $n$ scales very large.

    \item \textbf{Analysis of Adaptive Approaches (AFS vs SP-LSD):}
    AFS (137.8 ms) outperformed the BNRS (154.1 ms), leveraging data skew to skip rounds. However, it (AFS) trailed SP-LSD (106.0 ms) by a wide margin. Despite being `adaptive', AFS remains constrained by the predicted `Fat Bucket' problem and `cache inefficiency', thrashing associated with random-access partitioning. This suggests that while recursive MSD is viable, the overhead of its specific partitioning mechanism limits its potential compared to the linear scanning model of SP-LSD.

    \item \textbf{SP-LSD Dominance and Boundaries:}
    SP-LSD (106.0 ms) demonstrates a commanding performance advantage, establishing itself as the optimal choice. In relation to the competitors at $n=10^6$:
    \begin{itemize}
        \item It is 1.63x faster than Introsort, validating the efficiency of radix approaches for this domain.
        \item It is 1.45x faster than BNRS, proving that the overhead of stable partitioning is fully amortized by the reduction in sorting volume ($p$).
        \item It is 1.30x faster than AFS, confirming that the iterative linear scanning of SP-LSD is architecturally superior to the recursive MSD approach for fixed-width integers.
    \end{itemize}
    Surpassing Introsort, a hybrid-adaptive standard library algorithm with aggressive optimizations, using a core logical implementation suggests a significant algorithmic advantage for the target dataset. Unlike the approach of AFS, SP-LSD's fine-grained, iterative pruning model capitalized on data skew without incurring heavy structural overhead. Consequently, despite Introsort's efficiency on smaller arrays ($n \le 10,000$), SP-LSD is validated as the optimal choice for the target heavy-tailed dataset.
\end{enumerate}
While the `standard' adaptive approach (AFS) failed to outperform the baseline, SP-LSD bridged the gap to establish SP-LSD as the optimal algorithm for this dataset, proving that iterative binary pruning is the correct architectural choice for adaptive integer sorting.
    
\section{Conclusion} 
In this paper, we tackled the zero-padding inefficiency in Base-$n$ Radix Sort by developing the Stable Partitioned LSD (SP-LSD) model. Through the Radix Crossover Framework (RCF), we deterministically established the asymptotic, round-feasibility and pruning crossovers where SP-LSD outperforms both comparison and non-comparison sorting algorithms on both skewed and uniform distributions. The practical implementation of the work is benchmarked with Introsort, BNRS, AFS and from the results found that for the input size exceeding $10,000$, the proposed SP-LSD outperforms the competition in terms of execution time, repetitions, and overall operational cost. Future work will focus on extending the RCF to highly specialized high-performance systems, specifically exploring different optimization models targeting parallelization of the partition step and integrating within database query engines.


\bibliographystyle{splncs04}
\bibliography{BNRS-bib.bib}

@book{knuth98,
  author    = {Knuth, Donald E.},
  title     = {The Art of Computer Programming, Volume 3: Sorting and Searching (2nd Edition)},
  year      = {1998},
  isbn      = {0201896850},
  publisher = {Addison Wesley Longman Publishing Co., Inc.},
  address   = {USA}
}

@article{intro-sort97,
  author   = {Musser, David R.},
  title    = {Introspective Sorting and Selection Algorithms},
  journal  = {Software: Practice and Experience},
  volume   = {27},
  number   = {8},
  pages    = {983--993},
  year     = {1997}
}

@article{LAMARCA99,
  title    = {The Influence of Caches on the Performance of Sorting},
  journal  = {Journal of Algorithms},
  volume   = {31},
  number   = {1},
  pages    = {66--104},
  year     = {1999},
  issn     = {0196-6774},
  doi      = {10.1006/jagm.1998.0985},
  author   = {Anthony LaMarca and Richard E Ladner}
}

@article{ARS93,
  author    = {Peter M. McIlroy and Keith Bostic and M. Douglas McIlroy},
  title     = {Engineering Radix Sort},
  journal   = {Computing Systems},
  volume    = {6},
  number    = {1},
  pages     = {5--27},
  year      = {1993}
}

@article{Hoare62,
  author  = {Hoare, C. A. R.},
  title   = {Quicksort},
  journal = {The Computer Journal},
  volume  = {5},
  number  = {1},
  pages   = {10--16},
  year    = {1962}
}
\newpage

\section*{Appendix}
\begin{table}[h!]
\centering
\caption{Table of Notations Used in the Algorithm.}
\label{tab:algorithm_notations}
\begin{tabular}{|l|l|}
\hline
\textbf{Notation} & \textbf{Description}\\
\hline
$n$ & Total number of elements in the input array (i.e., input size). \\
$k$ & The maximum value for an element in the array. \\
$A$ & The input array containing $n$ elements \\
$A[i]$ & The element at index $i$ in the array $A$. \\
$B$ & An auxiliary output array for storing intermediate results. \\
$C$ & An auxiliary array used for counting frequencies. \\
$R$ & The number of rounds (passes), calculated as $\lfloor \log_n(k) \rfloor + 1$. \\
$r$ & The current round or iteration, where $0 \le r < R$. \\
$p$ & The proportion of elements removed by the first predicate check. \\
div & The value of an element scaled by the base, calculated as $\lfloor A[i] / \text{base}^r \rfloor$. \\
predicate & A check used to partition elements, e.g., $\lfloor A[i] / \text{div} \rfloor$. \\
$\alpha$ & Amortized cost of random-access sorting per element. \\
$\beta$ &  Amortized cost of partitioning per element.\\
$c$ & Ratio of sorting cost to partitioning cost.\\
\hline
\end{tabular}
\end{table}
\subsection{Time and Space Complexities of Sorting Algorithms}
\tscsort
\subsection{American Flag Sorting}
\AFS
\subsection{Comparison Sortings}
\comparsort

\subsection{Crossover Calculations for Different Datatypes}
\datatypecal
\subsection{Algorithms - SP-LSD and Demonstration of SP-LSD}
\Algosandex
\subsubsection{Demonstration of BNRS}
\demonstrationofBNRS


\end{document}